\documentclass[floatfix,preprint,preprintnumbers,superscriptaddress,amsmath,amssymb,pra]{revtex4-1}

\usepackage{graphicx}% Include figure files
\usepackage{bm}
\usepackage{color}

\begin{document}

\preprint{}

\title{Nodal superconductivity coexists with low-moment static magnetism in single-crystalline tetragonal FeS: A muon spin relaxation and rotation study}

\author{C. Tan}
\author{T. P. Ying}
\author{Z. F. Ding}
\author{J. Zhang}
\affiliation{State Key Laboratory of Surface Physics, Department of Physics, Fudan University, Shanghai 200433, China}
\author{D. E. MacLaughlin}
\affiliation{Department of Physics and Astronomy, University of California, Riverside, California 92521, USA}
\author{O. O. Bernal}
\affiliation{Department of Physics and Astronomy, California State University, Los Angeles, California 90032, USA}
\author{P. C. Ho}
\affiliation{Department of Physics, California State University, Fresno, California 93740, USA}
\author{K. Huang}
\affiliation{State Key Laboratory of Surface Physics, Department of Physics, Fudan University, Shanghai 200433, China}
\affiliation{National High Magnetic Field Laboratory, Tallahassee, Florida 32310, USA}
\author{I. Watanabe}
\affiliation{Advanced Meson Science Laboratory, RIKEN Nishina Center, Wako 351-0198, Japan}
\author{S. Y. Li}
\affiliation{State Key Laboratory of Surface Physics, Department of Physics, Fudan University, Shanghai 200433, China}
\affiliation{Collaborative Innovation Center of Advanced Microstructures, Nanjing 210093, China}
\author{L. Shu}
\thanks{Corresponding author: leishu@fudan.edu.cn}
\affiliation{State Key Laboratory of Surface Physics, Department of Physics, Fudan University, Shanghai 200433, China}
\affiliation{Collaborative Innovation Center of Advanced Microstructures, Nanjing 210093, China}

\date{\today}
\begin{abstract}
We report muon spin relaxation and rotation ($\mu$SR) measurements on hydrothermally-grown single crystals of the tetragonal superconductor~FeS, which help to clarify the controversial magnetic state and superconducting gap symmetry of this compound. $\mu$SR time spectra were obtained from 280~K down to 0.025~K in zero field (ZF) and applied fields up to 20~mT\@. In ZF the observed loss of initial asymmetry (signal amplitude) and increase of depolarization rate~$\Lambda_\mathrm{ZF}$ below 10~K indicate the onset of static magnetism, which coexists with superconductivity below $T_c$. Transverse-field $\mu$SR yields a muon depolarization rate $\sigma_\mathrm{sc} \propto \lambda_{ab}^{-2}$ that clearly shows a linear dependence at low temperature, consistent with nodal superconductivity. The $s{+}d$-wave model gives the best fit to the observed temperature and field dependencies. The normalized superfluid densities versus normalized temperature for different fields collapse onto the same curve, indicating the superconducting gap structure is independent of field. The $T{=}0$ in-plane penetration depth~$\lambda_{ab}$(0) = 198(3)~nm.
\end{abstract}

%\pacs{75.10.Hk, 05.10.Ln, 64.60.Cn, 11.15.Ha }% PACS, the Physics and Astronomy
 %Classification Scheme.

\maketitle

\section{introduction}

The discovery of superconducting La(O$_{1-x}$F$_{x}$)FeAs~\cite{Kamihara2008iron} has triggered extensive studies on iron-based superconductors (IBS)~\cite{Chen2014Iron,Dai2015Antiferromagnetic}. Most of the IBS share the same common structural motif of Fe-As layers, and the highest $T_c$ value is up to 56~K~\cite{Ren2008Superconductivity,Wu2009Superconductivity}. Density functional theory (DFT) calculations showed similarities of Fermi-surface structure between Fe-As based superconductors and iron chalcogenides (FeSe, FeS and FeTe)~\cite{Subedi2008Density}. Iron chalcogenides have the simplest crystal structure (iron chalcogenide layers) of IBS, and therefore have attracted great interest~\cite{Yoshikazu2010Review}. FeSe, the most studied iron chalcogenide, shows superconductivity below 8~K~\cite{hsu2008superconductivity}, relatively lower than iron arsenide superconductors. However, the superconducting transition temperature $T_c$ increases drastically under pressure~\cite{Medvedev2009Electronic}, by carrier doping~\cite{Miyata2015High}, or by growing single-layer FeSe on a SrTiO$_3$ substrate~\cite{Wang2012Interface,Ge2015Superconductivity}. Nematic order~\cite{FCS14} occurs in bulk FeSe below $T_s = 90$~K~\cite{McQueen2009Tetragonal}, and antiferromagnetic (AFM) order is absent~\cite{baek2015orbital,Bohmer2015Origin}. This makes FeSe a clean platform to study the nature of Fe-based superconductivity. However, its superconducting gap structure remains controversial~\cite{Khasanov2008Evidence,Dong2009Multigap,Song2011Direct}.

Recently, superconducting tetragonal FeS ($T_c \approx 4.5$~K) was successfully synthesized by Lai \textit{et al.}~\cite{Lai2015Observation} using a hydrothermal method. It has the same structure as FeSe, simply by replacing selenium with sulfur. Many studies have been made to understand the magnetic state and superconducting gap symmetry of FeS\@. Notably, two superconducting domes were observed under pressure~\cite{zhang2017observation}, posing challenges to understanding its pairing mechanism. Muon spin relaxation/rotation ($\mu$SR)~\cite{schenck1985muon,brewer1994encyclopedia,yaouanc2011muon} experiments on polycrystalline tetragonal FeS~\cite{Holenstein2016Coexistence,Kirschner2016Robustness} indicated fully-gapped superconductivity, and found low-moment disordered magnetism below $T_\mathrm{mag} \approx 20$~K~\cite{Holenstein2016Coexistence}. However, a nodal superconducting gap was observed in single-crystalline FeS by low temperature specific heat and thermal conductivity measurements~\cite{Xing2016Nodal,Ying2016Nodal}. Yang {\textit{et al.}}~\cite{Yang2016Electronic} calculated the electronic structure of FeS using DFT and reported that the gap function is nodal/nodeless on the hole/electron Fermi pockets. Soon after, angle-resolved photoemission spectroscopy (ARPES) studies~\cite{Miao2017Electronic} observed two hole-like and two electron-like Fermi pockets around the Brillouin zone center and corner, respectively. The authors attribute the controversies over the superconducting gap structure to the absence of a hole-like $\gamma$ band, which had been observed in other IBS\@. As for the magnetic properties, Man \textit{et al.}~\cite{Man2017Spin} concluded that FeS is a tetragonal paramagnet from elastic neutron scattering and transport measurements. This is consistent with a prediction of dynamical mean-field theory~\cite{Tresca2017Electronic}, but it contradicts the previous $\mu$SR results~\cite{Holenstein2016Coexistence}.

To help resolve these controversies we have performed $\mu$SR experiments on single crystals of tetragonal FeS\@. Our zero-field (ZF) and longitudinal-field (LF)-$\mu$SR measurements, made with  ensemble muon polarization (and applied field~$\mathbf{H}_L$ if present) parallel to the crystal \textbf{c} axis, revealed low-moment disordered static magnetism in the $ab$ plane below $T_\mathrm{mag} \approx 10$~K\@. Transverse field (TF)-$\mu$SR measurements in the superconducting state yield an in-plane penetration depth $\lambda_{ab}(0) = 198(4)$~nm, and reveal a linear temperature dependence as $T \to 0$, characteristic of an order parameter with line nodes. The temperature dependencies of the penetration depth measured at various applied fields are best described by a $s{+}d$-wave model. The normalized superfluid densities versus normalized temperature collapse onto a universal curve, indicating that the superconducting gap structure of FeS is field-independent.

\section{experiments and results}

Single-crystalline tetragonal FeS was prepared by de-intercalation of potassium cations from K$_\mathrm{x}$Fe$_{2-y}$S$_2$ ($x \approx 0.8$, $y \approx 0.4$) single crystals by hydrothermal reaction~\cite{Borg2016strong,Lin2016Multiband}. Elemental analysis, X-ray diffraction (XRD), scanning electron microscopy image, magnetic susceptibility and in-plane resistivity measurements were carried out, with results that are consistent with previously reported work~\cite{zhang2017observation,Ying2016Nodal}. Two batches of single crystals were prepared, denoted as Sample~A and Sample~B\@. $\mu$SR experiments were performed on the M15 and M20 spectrometers at TRIUMF, Vancouver, Canada. ZF- and LF-$\mu$SR measurements were carried out over the temperature range 25~mK--280~K for fields up to 20~mT\@. TF-$\mu$SR measurements were performed from 6~K down to 25~mK at various fields.

\subsection{ZF-$\bm{\mu}$SR} \label{sect:ZFuSR}

\begin{figure}[ht]
 \centering
 \includegraphics[width=0.48\textwidth]{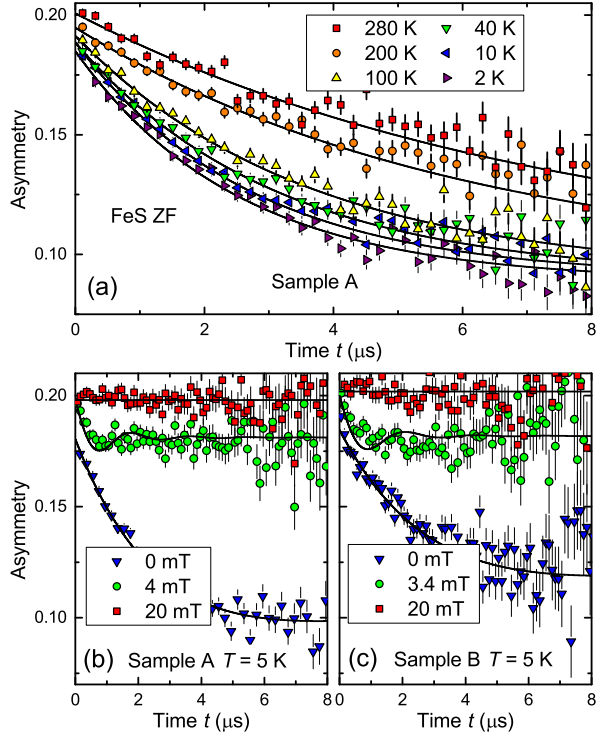}
 \caption{(Color online) (a) ZF-$\mu$SR spectra from single-crystalline tetragonal FeS at representative temperatures. Curves: fits to the data by a simple exponential decay function [Eq.~(\ref{eq.ZF})]. ZF- and LF-$\mu$SR time spectra for FeS (b) Sample~A and (c) Sample~B for various longitudinal fields~$H_L$ at 5~K\@. Curves: fits of the LF Lorentzian Kubo-Toyabe function~\cite{Uemura1985Muon,yaouanc2011muon} to the LF data.}
 \label{fig.1}
\end{figure}

The evolution in time of the decay positron count asymmetry, which is proportional to the muon depolarization, is often called a $\mu$SR spectrum. Representative ZF-$\mu$SR spectra between 2~K and 280~K are presented in Fig.~\ref{fig.1}%\hyperref[fig.1]
{(a)}.
The muon depolarization is well described by a simple exponential decay function
\begin{equation}
 \label{eq.ZF}
 A(t)/A_0 = (1-f)\exp(-\Lambda_\mathrm{ZF}t) + f
\end{equation}
at all temperatures. Here $A_0$ is the initial magnitude of the asymmetry signal and $\Lambda_\mathrm{ZF}$ is the ZF muon depolarization rate. The constant fraction~$f$ is the sum of two terms:
\begin{equation} \label{Eq:AgKT}
f = f_\mathrm{Ag} + f_\mathrm{ZF},
\end{equation}
where $f_\mathrm{Ag}$ is the fraction of muons that miss the sample and stop in the silver sample holder, and $f_\mathrm{ZF}$ is the fraction of local-field component parallel to the initial muon spin. This local-field component causes no precession and hence no depolarization in the absence of dynamic spin relaxation. For randomly-oriented local fields $f_\mathrm{ZF} = 1/3$, and for local fields perpendicular to the muon polarization $f_\mathrm{ZF} = 0$.

In ZF these two contributions cannot be distinguished. In TF-$\mu$SR, however, there is no analog to $f_\mathrm{ZF}$ in Eq.~(\ref{Eq:AgKT}). The observed values of $f$ and $f_\mathrm{Ag}$ obtained from TF-$\mu$SR data (Sec.~\ref{sec:TFuSR}) are nearly the same, i.e., $f_\mathrm{ZF} \approx 0$, consistent with internal fields at muon sites that are in the $ab$-plane. We note that the natural abundances and nuclear magnetic moments of both $^{57}$Fe and $^{33}$S are small~\cite{yaouanc2011muon}, and the Gaussian Kubo-Toyabe relaxation expected from their dipolar fields is negligible.

Simple exponential muon depolarization is usually caused either by motionally-narrowed dynamic relaxation, or a Lorentzian static field distribution~\cite{Uemura1985Muon,yaouanc2011muon}. The muon depolarization for a randomly-oriented static local field distribution is described by a static Kubo-Toyabe (KT) function~\cite{KuTo67,Hayano1979Zero}. An applied magnetic field~$\mu_0H_L \gg \Lambda_\mathrm{ZF}/\gamma_\mu \approx 0.5$~mT, where $\gamma_\mu = 851.616$~MHz/T is the muon gyromagnetic ratio, ``decouples'' the local field~\cite{Hayano1979Zero,Uemura1985Muon,yaouanc2011muon} (i.e., prevents muon precession). As shown in Figs.~\ref{fig.1}%\hyperref[fig.1]
{(b)} and %\ref{fig.1}\hyperref[fig.1]
{(c)}, at 5~K muon depolarization is completely suppressed in a field~$\mu_0H_\mathrm{L} = 20$~mT, indicating the local field is (quasi)static. The $\mu$SR spectra for intermediate fields can be fitted by the LF KT function appropriate to a Lorentzian static field distribution~\cite{Uemura1985Muon} although, as noted above, the local fields are unlikely to be randomly oriented.

\begin{figure}[ht]
 \centering
 \includegraphics[width=0.48\textwidth]{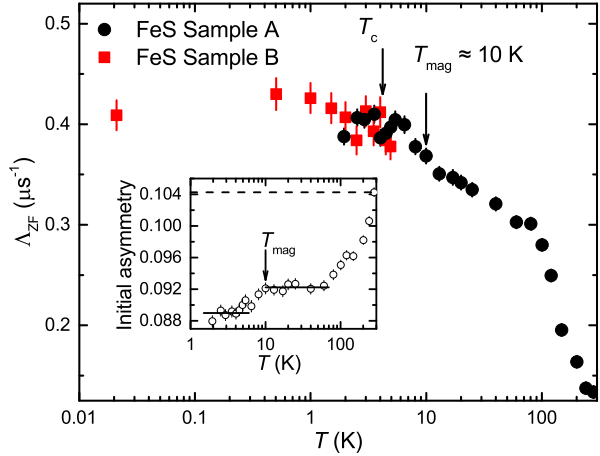}
 \caption{(Color online) ZF muon depolarization rate $\Lambda_\mathrm{ZF}$ versus logarithmic temperature. Inset: temperature dependence of the sample initial asymmetry. Dashed line: full asymmetry of the sample at 280~K\@. The solid curves are guides to the eye. The onset of static magnetism is evidenced by the increase of $\Lambda_\mathrm{ZF}$ and additional loss of initial asymmetry below $T_\mathrm{mag} \approx 10$~K.}
 \label{fig.2}
\end{figure}

The temperature dependencies of $\Lambda_\mathrm{ZF}$ and the sample initial asymmetry are given in Fig.~\ref{fig.2} and its inset, respectively.
The decrease of initial asymmetry with decreasing temperature above $\sim$80~K is due to the onset of a strong local field in a fraction of the sample volume, so that muons in this volume are rapidly depolarized and do not contribute to the signal~\cite{[{See, e.g., }] de1997muon}. This ``lost'' volume fraction increases with decreasing temperature, to 12\% at 80~K\@. Magnetic susceptibility and XRD measurements on our FeS single crystals show no signature of spurious impurity phases, indicating that their volume fraction is much less than 12\%. A similar loss of initial asymmetry was observed in ZF-$\mu$SR measurements on polycrystalline FeS samples~\cite{Holenstein2016Coexistence}, where it was attributed to small grains of a ferromagnetic impurity phase. These produce stray fields that affect an increasing fraction of the sample with decreasing temperature. This observation in both single-crystal and polycrystal FeS samples suggests that a spurious ferromagnetic phase is a byproduct of hydrothermally grown FeS~\cite{Holenstein2016Coexistence, Kirschner2016Robustness}.

The anomaly in $\Lambda_\mathrm{ZF}$(T) at 80~K (which was not reported in Ref.~\cite{Holenstein2016Coexistence}) is close to a structural transition temperature for FeSe~\cite{McQueen2009Tetragonal}, and is reminiscent of the possibility of nematic order~\cite{FCS14}. However, neither a structural transition nor nematic order has been observed in FeS~\cite{Man2017Spin,pachmayr2016structural}. The lattice parameters of tetragonal FeS decrease with decreasing temperature above 100~K, and remain almost constant below 100~K with a change of less than 1\% from the value at 300~K~\cite{pachmayr2016structural}. Excluding these possibilities, the increase of $\Lambda_\mathrm{ZF}$(T) with decreasing temperature above 80~K is most probably due to increased local fields as discussed above. This in turn suggests a distribution of impurity-phase Curie temperatures~$T_C$.

Between 10~K and 80~K, the initial asymmetry is temperature independent. This is consistent with the anomaly in $\Lambda_\mathrm{ZF}$ at 80~K, and suggests that 80~K is the minimum in the distribution of $T_C$; all impurity-phase grains are ferromagnetic below this temperature. The increase of $\Lambda_\mathrm{ZF}$ with decreasing temperature below 80~K is then probably intrinsic to FeS and dynamic, due to slowing down of intrinsic magnetic moment fluctuations. Future LF-$\mu$SR experiments will be necessary to determine separate static and dynamic contributions to $\Lambda_\mathrm{ZF}$ in this temperature range.

From 10~K to ${\sim}T_c$ the initial asymmetry decreases and $\Lambda_\mathrm{ZF}(T)$ increases further, indicating a second source of static magnetism with a distribution of ordering temperatures~\cite{de1997muon}. The absence of oscillations in ZF-$\mu$SR spectra [Fig.~\ref{fig.1}(a)] indicates that this static magnetism is disordered. The exponential form of the muon depolarization discussed in Sect.~\ref{sect:ZFuSR} is expected in dilute spin glasses~\cite{Uemura1985Muon}, where the required Lorentzian field distribution is a consequence of the $1/r^3$ spatial dependence of the dipolar local field, but a ``Lorentzian'' distribution can arise from aspects of the disorder other than dilution. Here the origin is probably low-moment short-range static magnetism~\cite{Holenstein2016Coexistence} with considerable inhomogeneity.

If we assume that the muon site in FeS is the same as calculated for isostructural FeSe~\cite{Bendele2012Coexistence}, then $\Lambda_\mathrm{ZF} \sim 0.4~\mu$s$^{-1}$ corresponds to a Fe magnetic moment of the order of $10^\mathrm{-3}~\mu_{B}$~\cite{Holenstein2016Coexistence}. Such a small moment would be undetectable by neutron diffraction. It should be noted, however, that the calculated muon stopping site~\cite{Bendele2012Coexistence} possesses a high point symmetry, so that partial cancellation of local fields is possible if the short-range correlation is AFM\@. The above estimate does not take this into account, so that the actual Fe magnetic moment could be considerably higher.

Below $T_c$ $\Lambda_\mathrm{ZF}$ saturates at ${\sim}0.42~\mu\mathrm{s}^{-1}$, and the initial asymmetry is again constant. The fact that exponential relaxation characterizes $\sim$85\% of the sample (Fig.~\ref{fig.2} inset) shows that the low-moment static magnetism coexists with superconductivity without the competition observed in other IBS~\cite{Drew2009Coexistence,Bendele2010Pressure,Shermadini2011Coexistence} where the volume fraction of magnetism is constant below $T_c$.

\subsection{TF-$\bm{\mu}$SR} \label{sec:TFuSR}

In a type-II superconductor an applied magnetic field can induce a flux line lattice (FLL), in which the distribution of the field is determined by the magnetic penetration depth~$\lambda$, the vortex core radius, and the structure of the FLL~\cite{Brandt1988Flux}. In a TF-$\mu$SR setup, a field is applied perpendicular to the initial muon spin polarization. The distribution of precession frequencies in a FLL and resulting loss of ensemble muon spin polarization reflect the field inhomogeneity, and quantities such as $\lambda$ can be extracted from the $\mu$SR spectra~\cite{Sonier2000muSR,yaouanc2011muon}.

For a perfect FLL the distribution of internal field is highly asymmetric, far from either a Gaussian or a Lorentzian field distribution. Weak random pinning slightly distorts the FLL so that the extrema of the field distribution fluctuate spatially; this often makes a Gaussian field distribution a good approximation ~\cite{Brandt1988Flux}. The muon spin depolarization rate $\sigma_\mathrm{sc}$ is related to the root-mean-square variation~$\Delta B_\mathrm{rms} = \overline{(\Delta B^2)}^\mathrm{1/2}$ of the internal field in the FLL\@. In turn, $\Delta B_\mathrm{rms}$ is proportional to $\lambda$, which is often estimated from the relation
\begin{equation}
 \label{eq.depth}
 \Delta B_\mathrm{rms}^2 =\sigma_\mathrm{sc}^2/\gamma_\mu^2 = 0.00371\Phi_0^2\lambda^{-4}
 \end{equation}
appropriate to an extreme type-II (London) superconductor with Ginzburg-Landau (GL) parameter~$\kappa = \lambda/\xi \gg 1$~\cite{Brandt2003Properties}. Here $\Phi_0$ = 2.068$\times$10$^{-15}$ Wb is the magnetic flux quantum.

\begin{figure}[ht]
 \centering
 \includegraphics[width=0.48\textwidth]{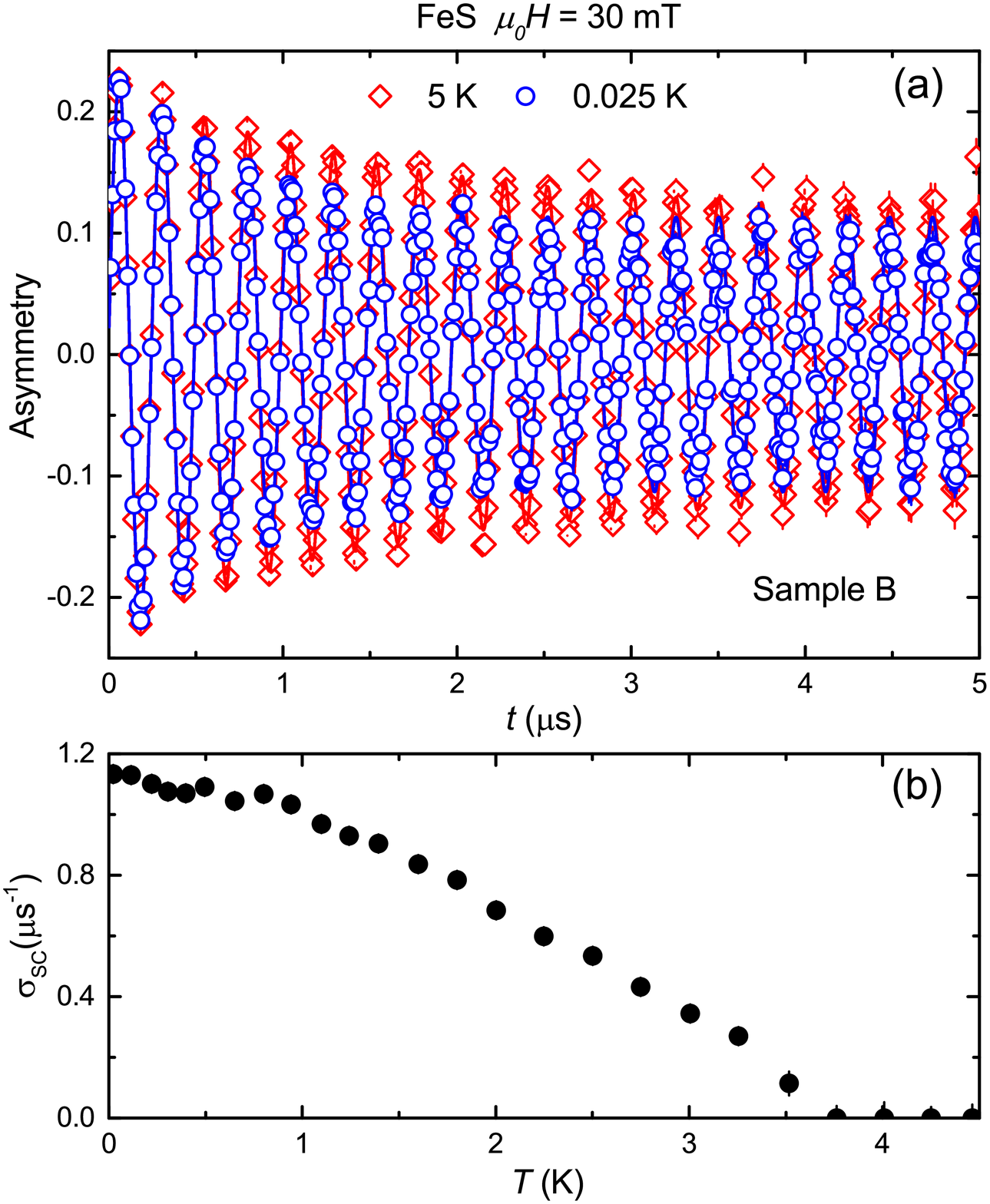}
 \caption{(Color online) (a) TF-$\mu$SR time spectra for FeS Sample~B above (squares) and below (circles) the superconducting transition temperature~$T_c = 4.1$~K\@. Solid lines: fits to the data by Eq.~(\ref{eq.TF}). The additional muon depolarization below $T_c$ is due to the field distribution in the FLL. (b) Temperature dependence of the Gaussian depolarization rate $\sigma_\mathrm{sc}$ from fits of Eq.~(\ref{eq.TF}) to TF-$\mu$SR data measured at $\mu_{0}H$ = 30~mT.}
 \label{fig.3}
\end{figure}

TF-$\mu$SR data were taken after cooling from the normal state in constant field, since changing the field below $T_c$ produces spurious field inhomogeneity due to flux trapping. Figure~\ref{fig.3}%\hyperref[fig.3]
{(a)} gives representative TF-$\mu$SR spectra for FeS Sample~B at $\mu_0${H} = 30~mT above and below $T_c$.
These spectra are well described by the TF muon depolarization function
\begin{equation}
 \label{eq.TF}
 \begin{split}
 A(t)/A_0 =\ & (1-f_\mathrm{Ag})\exp(-\Lambda_\mathrm{TF}t - {\textstyle\frac{1}{2}}\sigma_\mathrm{sc}^2 t^2)\cos(\gamma_\mu Bt+\varphi)\\
 & +\ f_\mathrm{Ag}\cos(\gamma_\mu B_\mathrm{ext} t+\varphi_\mathrm{Ag})\mathrm{,}
 \end{split}
\end{equation}
where $\Lambda_\mathrm{TF}$ is the depolarization rate due to static magnetism (in analogy to $\Lambda_\mathrm{ZF}$), $\sigma_\mathrm{sc}$ is the Gaussian depolarization rate due to the FLL, and $B$ and $\varphi$ are the mean field and initial phase of the ensemble muon precession, respectively.
The muon depolarization above $T_c$ is due only to static magnetism, and exhibits a simple exponential character (Fig.~\ref{fig.1}) similar to ZF data. Below $T_c$ $\Lambda_\mathrm{TF}$ is fixed to its value above $T_c$ ($\sim 0.63~\mu$s$^{-1}$ ), which is slightly larger than $\Lambda_\mathrm{ZF}$. This suggests that the applied field drives the in-plane local field slightly out of the plane.

Below $T_c$ a Gaussian muon depolarization is induced by the inhomogeneous field distribution due to the FLL\@. Fig.~\ref{fig.3}%\hyperref[fig.3]
{(b)} shows the temperature dependence of the Gaussian depolarization rate~$\sigma_\mathrm{sc}$.

The temperature dependence of $\sigma_\mathrm{sc}$, which is proportional to the superfluid density $\sigma_\mathrm{sc} \propto n_s \propto 1/\lambda_{ab}^2$, is fitted by the relation~\cite{Kim2002Reflection,Khasanov2007Experimental,Khasanov2008Evidence}:
\begin{equation}\label{Eq.Gap}
\frac{\sigma_\mathrm{sc}(T)}{\sigma_\mathrm{sc}(0)} = 1+\frac{1}{\pi}\int_0^{2\pi}\!\!\!\int_{\Delta(T,\varphi)}^{\infty} \!\!\!dE\,d\varphi\,\frac{\partial f}{\partial E}\frac{E}{\sqrt{E^2-\Delta^2(T,\varphi)}},
\end{equation}
where $f(E)$ is the Fermi function. The gap symmetry enters this expression via the form of $\Delta_s(T,\varphi)$. For the $s$-wave model $\Delta_s(T,\varphi) = \Delta^s(0)\delta(T/T_c)$, where the temperature dependence~$\delta(T/T_c)$ of the superconducting gap is estimated using~\cite{Khasanov2007Experimental,Khasanov2008Evidence}
\begin{equation}
\delta(T/T_c) = \tanh\{1.82 [1.018(T_c/T-1)]^{0.51}\}.
\end{equation}
For the $d$-wave model, $\Delta_d(T,\varphi) = \Delta^d(0)\delta(T/T_c) \mathrm{cos}(2\varphi)$. In the recently-proposed orbital-selective $s\tau_3$ state for iron selenides~\cite{Nica2017Orbital}, the intraband ($d_{{x}^2-{y}^2}$) and interband ($d_{xy}$) nodal pairing terms add in quadrature. As a consequence, the quasiparticle excitation is fully gapped on the Fermi surface. A simplified model of the $s\tau_3$ state gives $\Delta_{s\tau_3}(T,\varphi) = \delta(T/T_c) [(\Delta_1(0)\mathrm{cos}(2\varphi))^2+(\Delta_2(0)\mathrm{sin}(2\varphi))^2]^{1/2}$~\cite{pang2016evidence}. Finally, for two weakly coupled superconducting bands (i.e., an $s$ band and a $d$ band), a linear combination of terms of the form of  Eq.~(\ref{Eq.Gap}) can be applied~\cite{Khasanov2007Experimental,Kim2002Reflection}:
\begin{equation}
 \label{eq.Sumgap}
 \begin{split}
 \frac{\sigma_\mathrm{sc}(T)}{\sigma_\mathrm{sc}(0)} =~&~\frac{\lambda^{-2}(T)}{\lambda^{-2}(0)} \\
 =~&~\omega\frac{\lambda^{-2}(T,\Delta_1(T))}{\lambda^{-2}(0,\Delta_1(0))}
 +(1-\omega)\frac{\lambda^{-2}(T,\Delta_2(T))}{\lambda^{-2}(0,\Delta_2(0))}.
 \end{split}
\end{equation}

\begin{figure*} [ht]
 \centering
 \includegraphics[width=\textwidth]{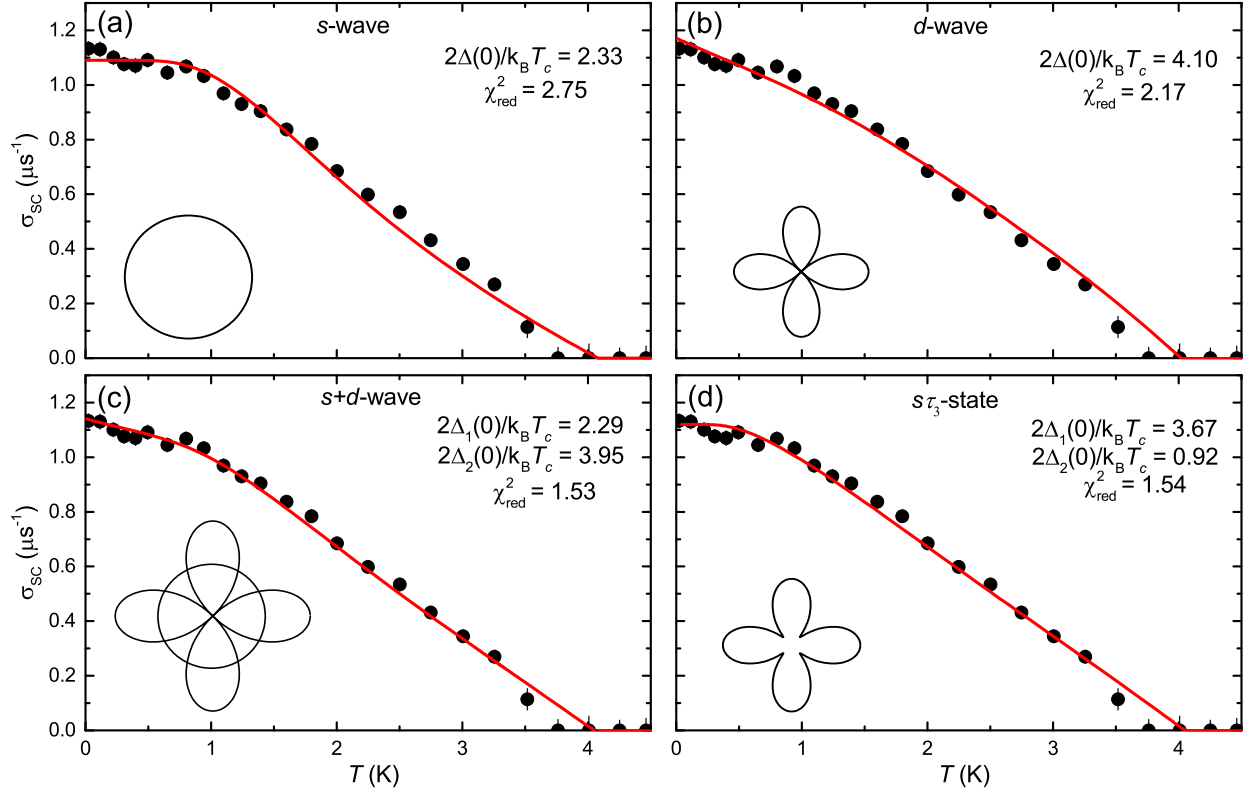}
 \caption{(Color online) Temperature dependence of the Gaussian depolarization rate $\sigma_\mathrm{sc}$ for FeS Sample~B. The solid lines are fits for different superconducting gap symmetries, shown schematically: (a) $s$-wave, (b) $d$-wave, (c) $s{+}d$-wave and (d) $s\tau_3$-state. The corresponding angular dependence of superconducting energy gap(s) are shown in insets. See main text for details.}
 \label{fig.Fit}
\end{figure*}

Fits of $s$-wave, $d$-wave, $s{+}d$-wave, and orbital-selective $s\tau_3$ models to our data are shown in Fig.~\ref{fig.Fit}.
The angular dependencies of the gaps are shown schematically in the insets. We use the reduced chi-square~$\chi^2_\mathrm{red}$ of the fits to evaluate the goodness of fit~\cite{Khasanov2008Evidence}. It is obvious that the single $s$-wave and $d$-wave models do not describe the temperature dependence of $\sigma_\mathrm{sc}$ accurately, and both $s{+}d$-wave and $s\tau_3$ models describe the data well. However, the $s{+}d$-wave model gives a better description of the low-temperature data. Fit parameters and $\chi^2_\mathrm{red}$ for these two models are shown in Table~\ref{Table.gap}.
\begin{table*} [ht]
\caption{Parameters from fits of the the $s{+}d$-wave and $s\tau_3$ models to the temperature dependence of $\sigma_\mathrm{sc}$.} \label{Table.gap}
 \begin{ruledtabular}
 \begin{tabular}{ccccccccccc}
 Model & $H$ (mT) & $T_c$ (K) & $\Delta_1$(0) (meV) & $\omega$ & $2\Delta_1/k_\mathrm{B}T_c$ & $\Delta_2$(0) (meV) & 1-$\omega$ & $2\Delta_2/k_\mathrm{B}T_c$ & $\chi^2_\mathrm{red}$ \\
 \hline
 $s{+}d$-wave         & 7.5  & 4.21 & 0.48  & 0.50  & 2.64 & 0.71  & 0.50 & 3.91 & 1.05 \\
 $s{+}d$-wave         & 30   & 4.05 & 0.4   & 0.43  & 2.29 & 0.69  & 0.57 & 3.95 & 1.53 \\
 $s{+}d$-wave         & 75   & 3.62 & 0.37  & 0.47  & 2.37 & 0.64  & 0.53 & 4.10 & 1.95 \\
 $s\tau_3$-state & 7.5  & 4.15 & 0.17  &     & 0.95 & 0.71  &    & 3.97 & 1.15 \\
 $s\tau_3$-state & 30   & 4.05 & 0.16  &     & 0.92 & 0.64  &    & 3.67 & 1.54 \\
 $s\tau_3$-state & 75   & 3.58 & 0.14  &     & 0.91 & 0.58  &    & 3.76 & 1.81 \\
 \end{tabular}
 \end{ruledtabular}
\end{table*}

Thus our results suggest an $s{+}d$-wave pairing state with multi-band and nodal superconductivity. Table~\ref{Table.gap} shows that the $s$ band and the $d$ band make comparable contributions to the superfluid density ($\omega \approx 0.5$), which is consistent with similar $\chi^2_\mathrm{red}$ from fits by single $s$-wave and $d$-wave models [Figs.~\ref{fig.Fit}%\hyperref[fig.Fit]
{(a)} and \ref{fig.Fit}%\hyperref[fig.Fit]
{(b)}]. Table~\ref{Table.gap} also shows that $2\Delta/k_BT_c$ for one gap is less than BCS value of 3.54 and is larger for the other gap. This is consistent with the theoretical constraints~\cite{Saito2013Nodal}, and has been observed in many IBS as summarized by Adamski {\textit{et al.}}~\cite{Adamski2017Signature}.

For some high-$T_c$ cuprates such as La$_\mathrm{1.83}$Sr$_\mathrm{0.17}$CuO$_\mathrm{4}$, the ratio of superfluid densities from different bands is very sensitive to the external field~\cite{Khasanov2007Experimental}. To investigate the field dependence of superconducting properties of tetragonal FeS, and also to give a better estimation of the absolute value of $\lambda$, we performed TF-$\mu$SR measurements at a number of applied fields.

\begin{figure}[ht]
 \centering
 \includegraphics[width=0.48\textwidth]{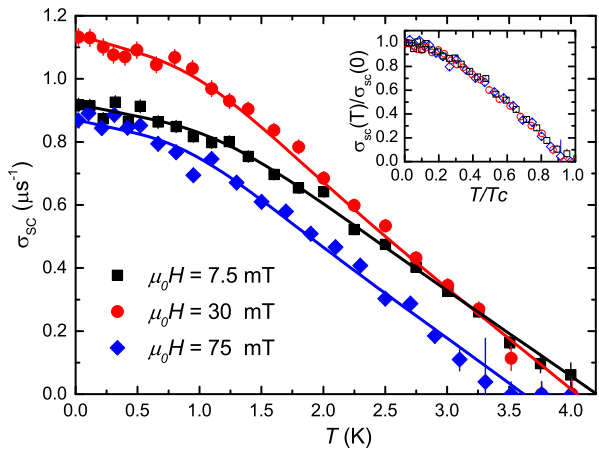}
 \caption{(Color online) Temperature dependence of $\sigma_\mathrm{sc}$ measured at $\mu_0$H = 7.5mT, 30~mT, and 75~mT for $T < T_c$ in single crystal FeS Sample~B. Solid curves: fits by $s{+}d$-wave model. Inset: normalized superfluid density versus normalized temperature. }
 \label{fig.5}
\end{figure}

Figure~\ref{fig.5} gives the temperature dependence of $\sigma_\mathrm{sc} \propto \lambda_{ab}^{-2}$ for $\mu_0${H} = 7.5~mT, 30~mT, and 75~mT\@.
The theoretical curves are from the $s{+}d$-wave model, with fitting parameters shown in Table~\ref{Table.gap} (the superconducting gaps and $T_c$ are free parameters in all fits). The ratio~$\omega$ of $s$-wave to $d$-wave contribution is almost independent of field. Therefore the normalized superfluid densities $\lambda_{ab}^{-2}(T)/\lambda_{ab}^{-2}(0) = \sigma_\mathrm{sc}(T)/\sigma_\mathrm{sc}(0)$ versus normalized temperature $T/T_c$ collapse onto a universal curve, indicating that the superconducting gap structure is independent of field in FeS\@. The values of $2\Delta\mathrm{(0)}/k_BT_c$ are close for different fields as shown in Table \ref{Table.gap}, indicating the self-consistency of the fit.

As shown in Fig.~\ref{fig.5}, $\sigma_\mathrm{sc}(0)$ has a maximum value at $\mu_0H \approx 30$~mT, consistent with the properties of the ideal GL vortex lattice~\cite{Brandt2003Properties}. From the value of $\sigma_{sc}(0)$ at $\mu_0H$ = 30~mT and Eq.~(\ref{eq.depth}), we estimate the in-plane penetration depth $\lambda_{ab}(0) = 307(4)$~nm. However, this approximation is good only for large $\kappa \gtrsim 70$, and only for $\mu_0H$ near the maximum of $\sigma_\mathrm{sc}(T{=}0,H)$~\cite{Brandt2003Properties}. To obtain a better estimation we use the Abrikosov solution of the linearized GL theory~\cite{Brandt2003Properties}, which yields
\begin{equation}\label{eq.6}
 \sigma_\mathrm{sc}^2/\gamma_\mu^2 = 7.52 \times 10^{-4} \frac{\kappa^\mathrm{4}(1-b)^2}{(\kappa^2-0.069)^2}\frac{\Phi_0^2}{\lambda_{ab}^\mathrm{-4}};
\end{equation}
here $b = B/B_\mathrm{c2} \approx H/H_\mathrm{c2}$ is the normalized field. The upper critical field~$B_{c2} = \Phi_0/2\pi\xi^2$ for FeS is $\approx 0.4$~T for $\mathbf{H} \parallel \mathbf{c}$~\cite{Ying2016Nodal,Lin2016Multiband}. This gives a more accurate value~$\lambda_{ab}(0) = 198(4)$~nm, with a resulting $\kappa = \lambda/\xi \approx 11$.

\section{discussion}

Previous $\mu$SR experiments on polycrystalline FeS by Holenstein \textit{et al.}~\cite{Holenstein2016Coexistence} revealed a low-moment magnetism below $T_\mathrm{mag} \approx 20$~K, whereas no intrinsic static magnetism was observed by other $\mu$SR experiments~\cite{Kirschner2016Robustness} or by neutron scattering or transport experiments~\cite{Man2017Spin}. Our ZF-$\mu$SR experiments on single crystalline FeS confirm the onset of low-moment static magnetism in the $ab$ plane below a lower $T_\mathrm{mag} \approx 10$~K, which coexists with superconductivity below $T_c$. The present results and those of Ref.~\cite{Holenstein2016Coexistence} for the temperature dependencies of $\Lambda_\mathrm{ZF}$ and the initial asymmetry are more or less consistent, although Holenstein \textit{et al.}\ do not report an anomaly at 80~K\@.

We note that the muon depolarization functions are different between our ZF-$\mu$SR spectra and those of Ref.~\cite{Holenstein2016Coexistence}: these authors report ``root exponential'' relaxation~$\exp[-(\Lambda t)^{1/2}]$, whereas we observe simple exponential relaxation. The difference is consistent with our conclusion that the low-moment static magnetism is in the $ab$-plane as in FeSe~\cite{Bendele2012Coexistence}, since then the local field would be more disordered in randomly-oriented polycrystalline samples. The root exponential function, which signals a broad distribution of exponential rates~\cite{John06} (i.e., even broader than a Lorentzian distribution), would then be a better description for ZF-$\mu$SR spectra of polycrystalline FeS.

The ZF-$\mu$SR study of polycrystalline FeS by Kirschner \textit{et al.}~\cite{Kirschner2016Robustness} used a sum of two simple exponential functions to describe the muon depolarization. The authors reported a slow relaxation in 85\% volume fraction, attributed to intrinsic magnetic moments of the iron in FeS, and a fast relaxation with 15\% volume fraction attributed to a magnetic impurity phase. The difference between this result and the root-exponential relaxation reported in Ref.~\cite{Holenstein2016Coexistence} may not be primarily in the data, but instead a consequence of the fact that a fit to data of a relaxation function that is a sum of exponentials often does not determine the coefficients in the sum (or the distribution function in an integral) well; the problem is ill-conditioned~\cite{[{See, e.g., }] PTVF92}. A two-exponential function is difficult to distinguish from a ``stretched exponential''~$\exp[-(\Lambda t)^\alpha]\ (\alpha < 1)$ unless the two amplitudes are comparable and the rates are very different.

Our TF-$\mu$SR measurements suggest the $s{+}d$-wave superconducting pairing symmetry, demonstrating a nodal and multi-band superconductivity, which is different from the previous $\mu$SR results~\cite{Holenstein2016Coexistence,Kirschner2016Robustness}. A similar situation arose in early TF-$\mu$SR penetration depth measurements on high-$T_c$ cuprates~YBa$_2$Cu$_3$O$_{7-\delta}$ (YBCO). Experiments on polycrystal materials as well as first available single crystals indicated an isotropic $s$-wave order parameter~\cite{Harshman1989Magnetic,Pumpin1990Muon}. Nodal superconductivity was observed only after experiments on good single-crystalline YBCO showing a linear low temperature dependence of penetration depth~\cite{Hardy1993Precision}. The origin of the controversy in this case was that the in-plane penetration depth $\lambda_{ab}$ was estimated by measuring $\lambda_\mathrm{eff}$ of polycrystalline samples, assuming that the temperature dependence of penetration depth is isotropic along different crystal orientations~\cite{fesenko1991analytical}. However, experimental results showed that the temperature dependence of $\lambda_c$ is significantly different from that of $\lambda_a$ and $\lambda_b$ in YBCO~\cite{Khasanov2007Multiple}, i.e., the superconducting gap symmetry is different for different crystal orientations. The difference between $\mu$SR results for the gap symmetry in FeS might have the same origin, and could be resolved by measuring $\lambda_c(T)$ in single crystals.

The fits of $\sigma_\mathrm{sc}$(T) suggest the presence of weakly-coupled $s$-wave (nodeless) and  $d$-wave (nodal) bands, consistent with other results. ARPES measurements~\cite{Miao2017Electronic} observed two hole-like and two electron-like Fermi pockets around the Brillouin zone center and corner, respectively, where the gap function is nodal/nodeless on the hole/electron Fermi pockets~\cite{Yang2016Electronic}. Scanning tunneling microscopy (STM) experiments~\cite{Yang2016Strong} showed a V-shaped spectrum, which is best described by both anisotropic $s$-wave and s+$d$-wave model. The weight factor and energy gaps of $s{+}d$-wave model fit for the STM spectra are close to our fitting results. Nodal gap behavior is also inferred from low temperature heat capacity and thermal conductivity measurements~\cite{Xing2016Nodal,Ying2016Nodal}.

\section{conclusions}

In summary, we have studied the magnetic and superconducting properties of FeS single crystal samples by $\mu$SR\@. Low-moment in-plane disordered static magnetism is found below $T_\mathrm{mag} \approx 10$~K\@. A significant $T$-linear dependence of the in-plane penetration depth is observed at low temperatures, indicating a nodal superconducting gap. The temperature dependencies of the superfluid density are best described by the multi-band and nodal superconductivity of the $s{+}d$-wave model. The normalized temperature dependencies of normalized superfluid density collapse on a universal curve for different fields, suggesting that the superconducting gap structure is field independent. The absolute value of the in-plane $T{=}0$ penetration depth is estimated to be 198(4)~nm.

\begin{acknowledgments}

We are grateful to G. D. Morris, B. Hitti, and D. Arsenau of the TRIUMF CMMS for assistance during the experiments. This research is supported by the National Key Research and Development Program of China (Nos.~2017YFA0303104 and 2016YFA0300503), and the National Natural Science Foundation of China under grant nos.~11474060 and 11774061. Work at CSULA was funded by the U.S. NSF DMR/PREM-1523588. Research at CSU Fresno was supported by NSF DMR-1506677. Research at U. C. Riverside was supported by the UC Riverside Academic Senate.
\end{acknowledgments}

%\bibliographystyle{prsty}
%\bibliography{FeS-v17}
%

%\end{CJK*}
\end{document}